\documentclass[twocolumn,showpacs,amsmath,amssymb,aps,prl]{revtex4}   %%  4-pages !!
\usepackage{graphicx}% Include figure files
\usepackage{dcolumn}% Align table columns on decimal point
\usepackage{bm}% bold math
\usepackage{color}% color text
\usepackage{amsmath}

\renewcommand{\vec}[1]{\mathbf{#1}}

\newif\ifgraph

\graphtrue             %
\begin{document}
\title{
A quorum sensing active matter in a confined geometry}

\author{Yuxin Zhou$^{1}$, Yunyun Li$^{1}$}
\email{yunyunli@tongji.edu.cn}
\author{Fabio Marchesoni$^{1,2}$}
\email{fabio.marchesoni@pg.infn.it}
 \affiliation{$^{1}$ Center for Phononics and Thermal Energy Science, Shanghai
 Key Laboratory of Special Artificial Microstructure Materials and Technology,
 School of Physics Science and Engineering, Tongji University, Shanghai 200092, China}
 \affiliation{$^{2}$ Dipartimento di Fisica, Universit\`{a} di Camerino, I-62032 Camerino, Italy}

\date{\today}

\begin{abstract}
Inspired by the problem of biofilm growth, we numerically investigate clustering in a two-dimensional suspension of
active (Janus) particles of finite size confined in a circular cavity. Their dynamics is regulated
by a non-reciprocal mechanism that causes them to switch from active to passive above a certain
threshold of the perceived near-neighbor density ({\em quorum sensing}).
A variety of cluster phases -- glassy, solid (hexatic) and liquid -- is observed depending on the
particle dynamics at the boundary, the quorum sensing range, and the level of noise.

\end{abstract}
\maketitle

\section {Introduction}\label{intro}

Bacteria are capable of adjusting their motility to form large colonies, like
biofilms. While motile bacteria have the advantage to swim efficiently
towards food sources, biofilms aggregates are able to resist environmental
threats such as antibacterial substances. Understanding the basic physical
mechanisms of biofilm growth is a topic of ongoing
research by many teams worldwide. Recent studies suggest that a
motility-based clustering phenomenon is involved in the formation of bacterial swarms \cite{Mazza} and in the
transition from
bacterial swarms to biofilms \cite{Grobas2021}. Moreover, it is demonstrated that synthetic
active materials, such as Janus colloids, can undergo
motility-induced aggregation, not only via high-density steric mechanisms \cite{Fily}, or lower density mutual interactions \cite{Bizonne}, but also by simply adjusting their velocity according to the
direction \cite{Lavergne2019} and the local density of their peers
\cite{Bauerle2018}, largely insensitive to pair interactions. These situations have been modeled theoretically
using both particle-based models and field
theoretical approaches \cite{Cates2015}. In this context, it was shown that
the active systems may exhibit motility-induced phase separation (MIPS),
whereby self-propelled particles, with only repulsive interactions, form
aggregates by reducing their swimming speed in response to a local density
value greater than a given threshold (a mechanism called {\em quorum sensing}
\cite{QS1,QS2}).

In its simplest form, MIPS has been shown to be analogous to a
gas-liquid phase separation. However, recent non-equilibrium field
theories have predicted intriguing behaviors, like microphase
separation \cite{Tjhung2018} and an active foam phase with slowly
coalescing bubbles \cite{Fausti2021}.  In fact, our understanding of how the
microscopic details of the single-particle dynamics lead to different
collective behaviors is presently far from satisfactory. Finally, it has been shown
that motile {\em E. coli} bacteria spontaneously aggregate within minutes
when subject to controlled convective flows produced by a microfluidic device
\cite{Yadzi2012}. It is still unclear, however, which physical ingredients
are required for a minimal active-particles model to
reproduce such a behavior \cite{TU1,TU2}.

To a closer look, it is apparent that, while the emergence of steady
aggregates of motile particles is largely driven by the nature of their
mutual interactions, which ultimately influence their motility, the
properties of such aggregates are strongly determined by the combined action
of spatial confinement and fluctuations of both the suspension fluid and
the self-propulsion mechanism. In this Letter we revisit the model of
non-reciprocal particle interaction proposed by Bechinger and coworkers
\cite{Bauerle2018} (see also Ref. \cite{Lavergne2019}), by investigating the
effects of the particle dynamics against the container walls at different noise
levels. Contact and far-field reciprocal (pair) interactions have been
neglected; no alignment mechanism has been invoked to trigger particle
aggregation: Quorum sensing under spatial confinement is the one mechanism
considered here. As a result, we observed a variety of cluster
phases, glassy, solid (hexatic) and liquid, and determined the relevant model phase
diagram.

\section {Model}\label{model}

{\em Single particle dynamics}. We considered the simplest realization of
a synthetic microswimmer, namely a two-dimensional (2D) Janus particle (JP) \cite{JP}.
An active JP of label $i$ gets a
continuous push from the suspension fluid, which in
the overdamped regime amounts to a self-propulsion velocity, ${\vec v_0}_i$, with
constant modulus, $v_0$, and orientation, $\theta_i$, fluctuating with time
constant, $\tau_\theta$, under the combined action of thermal noise and
the rotational fluctuations intrinsic to the specific self-propulsion
mechanism.  In 2D. its bulk dynamics obeys the Langevin
equations \cite{TU3}
\begin{figure}[tp]
\centering \includegraphics[width=7cm]{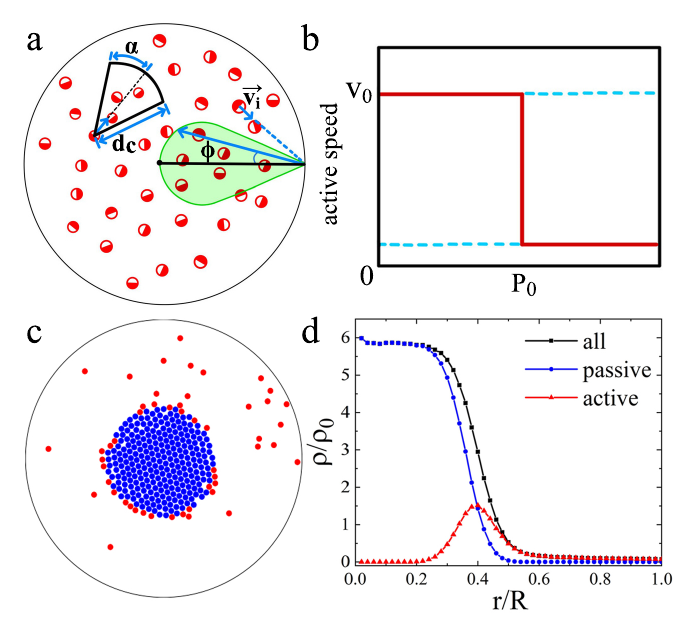}
\caption{Schematics of our model. (a) Illustration of the non-reciprocal
interaction mechanism for $N$ active Janus particles with visual half-angle $\alpha$ and horizon $d_c$. The
distribution, $p(\phi)$,  of the boundary scattering angle, Eq. (\ref{phi}), is plotted for
$\lambda= 100$: the distance of a point on the chart line from the scattering point on the boundary is proportional to the probability that the particle gets scattered in the direction of that point. (b) Quorum sensing protocol: for values of the sensing function above the threshold $P_{0}(\alpha)$, Eq. (\ref{Pth}), the particle turns from active to passive. (c) Example of passive cluster formation for $\lambda=100$, $\alpha=(7/8)\pi$, $d_c=16$, $D_\theta=0.001$, $v_0=0.5$, $R=45$, $r_0=1$, $N=304$, and $D_0=0.01$ (snapshot taken at $t=2\cdot 10^4$). Active/passive particles are represented by red/blue circles. (d) Passive (blue), active (red) and total (black) radial particle distributions for the parameters in (c). These distributions have been averaged over time (10 snapshots taken every 1,000 time units starting at $t=10^4$) and initial conditions (200 realizations).\label{F1}}
\end{figure}
\begin{eqnarray}
\label{LE} \dot x_i &=& v_0\cos \theta_i +\xi_{xi}(t) \\ \nonumber \dot y_i
&=& v_0\sin \theta_i +\xi_{yi}(t) \\ \nonumber \dot \theta_i &=&\xi_{\theta i}(t),
\end{eqnarray}
where ${\bf r}_i=(x_i,y_i)$ are the coordinates of the particle center of mass
subject to the Gaussian noises $\xi_{pi}(t)$, with $\langle
\xi_{qi}(t)\rangle=0$ and $\langle
\xi_{qi}(t)\xi_{pi}(0)\rangle=2D_0\delta_{qp}\delta (t)$ for $q,p=x,y$, modeling
the equilibrium thermal fluctuations in the suspension fluid.  The
orientational fluctuations of the propulsion velocity are modeled by the
Gaussian noise $\xi_{\theta i}(t)$ with $\langle \xi_{\theta i}(t)\rangle=0$ and
$\langle \xi_{\theta i}(t)\xi_{\theta i
}(0)\rangle=2D_{\theta}\delta(t)$, where
$D_{\theta}=1/\tau_{\theta}$ is the relaxation rate of the self-propulsion
velocity.

The simplifications introduced in Eq. (\ref{LE}) are not limited to the
reduced dimensionality of the system. All noise sources have been treated as
independent, although, strictly speaking, spatial and orientational
fluctuations are statistically correlated to some degree. Moreover, we ignored
hydrodynamic effects  which may favor the clustering of active particles
at high packing fractions. However, we made sure that the parameters used in our
simulations are experimentally accessible, as apparent on expressing times in
seconds and lengths in microns. The stochastic differential Eqs. (\ref{LE})
were numerically integrated by means of a standard Euler-Maruyama scheme
\cite{Kloeden}. To ensure numerical stability, the numerical integrations
have been performed using an appropriately short time step, $10^{-3}$.
\begin{figure}[tp]
\centering \includegraphics[width=6.5cm]{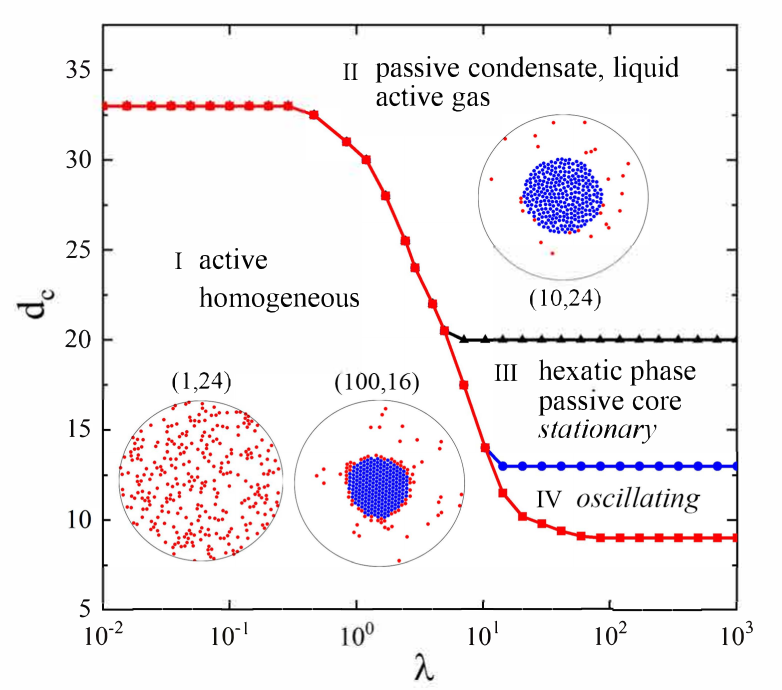}
\caption{Phase diagram in the space parameter $(\lambda, d_c)$. Four distinct regions are distinguishable, namely, I (below the red curve): all particles remain active; II: a liquid condensate of passive particle coexists with a gas of active particles; III: the passive condensate solidifies into a steady-state hexatic structure; and IV: periodic formation of hexatic passive clusters (see also Fig. \ref{F3}). Snapshots of the relevant suspension patterns taken at time $t=2\cdot 10^4$ are shown as insets for different $(\lambda, d_c)$; red and blue circles denote respectively active and passive
 particles. \label{F2}}
\end{figure}

{\em Boundary conditions}. In this study, the JPs are confined to a restricted
area, say, a circular cavity of radius $R$ [Fig. \ref{F1}(a)]. One can think
of motile bacteria spreading on a Petri dish. Equations (\ref{LE}) still
apply away from the walls; we only need to set a prescription to treat the
particle collisions with the boundaries.
Following Refs. \cite{Kaergar,Codina}, we assume that, upon hitting it, a JP is
captured by the wall and immediately re-injected into the cavity, at
an angle $\phi$ with respect to the radius (i.e., the
perpendicular) through the collision point [Fig. \ref{F1}(a)]. A finite (short)
trapping time does not affect the conclusions of the present work. A commonly
accepted distribution for such a boundary scattering angle is \cite{Kaergar}
\begin{eqnarray}
\label{phi} p(\phi)= 2\exp(\lambda \cos \phi)/[\pi I_0(\lambda)]
\end{eqnarray}
where $I_0$ is the modified Bessel function of the first kind and the parameter
$\lambda$ depends on the temperature and the physio-chemical
properties of the particle and the cavity wall.
Notice that in the limit $\lambda \to \infty$ we recover
the reflecting boundary conditions adopted in
Ref. \cite{Bauerle2018}, namely, $p(\phi)=\delta(\phi)$. We
considered other cavity geometries as well; an example is discussed at the
bottom of the forthcoming section.

{\em Non-reciprocal interaction (sensing)}.
When $N$ identical, independent  active particles of Eq. (\ref{LE}) are confined into the
cavity, interactions among them cannot be neglected. In our simulations we
consider only two kinds of interactions: {\em (i) hard-core repulsion},
whereby the particles are modeled as hard discs of radius $r_0$. Further
reciprocal interactions have been discarded; {\em (ii) neighbor
perception}, a mechanism governing the motility of each particle depending on
the spatial distribution of its neighbors. In biological systems this process
is mediated by some form of inter-particle communication (mostly chemical in
bacteria colonies \cite{QS1,QS2}). On the other hand, the motility of
artificial microswimmers grows less efficient with increasing their density
\cite{JP}. Without entering the details of the specific perception
mechanisms, we can define the sensing function of particle $i$
as follows \cite{Bauerle2018} (see also \cite{Patelli})
\begin{eqnarray}
\label{Pa} P_i(\alpha)= \sum_{j\in V_i^\alpha}\frac{1}{2\pi r_{ij}},
\end{eqnarray}
where $r_{ij}$ is the distance between particles $i$ and $j$ and $V_i^\alpha$
denotes the visual cone of particle $i$, centered around the direction of its
self-propulsion velocity, ${\vec v_0}_i$, with finite horizon, $r_{ij}\leq
d_c$. This means that each particle senses the presence of other particles
only within a restricted visual cone and a finite distance, $d_c$. For a
uniform active suspension of density, $\rho_0=N/\pi R^2$, the sensing function
of a particle placed at the center of the cavity reads \cite{Lavergne2019}
\begin{eqnarray}
\label{Pth} P_0(\alpha)= ({\alpha}/{\pi})\rho_0 R
\end{eqnarray}
We assume now that the particle motility is governed by the following simple
{\em quorum sensing protocol} [Fig. \ref{F1}(b)],
\begin{eqnarray}
\label{QS-e}
|{\vec v_0}_i|=
\begin{cases}
v_{0}& P_i(\alpha) \leq P_0(\alpha) \\
0 & P_i(\alpha) > P_0(\alpha).
\end{cases}
\end{eqnarray}
Clearly,  this
form of particle interaction is non-reciprocal, since $j$ may be perceived by
$i$ and, therefore, influence its dynamics, without being itself affected by
the presence of $i$. The dynamical implications of the non-reciprocal interactions in
biological matter are discussed at length by Bechinger and coworkers in
Refs. \cite{Lavergne2019,Bauerle2018}. For an earlier and more
elaborated quorum sensing model of synthetic active matter,
the reader is referred to Ref. \cite{Barberis}. What matters here, is that
for appropriate choices of the horizon range, $d_c$, and the
visual angle, $\alpha$, clustering may occur, as illustrated in Fig. \ref{F1}(c).

\begin{figure}[tp]
\centering \includegraphics[width=6.5cm]{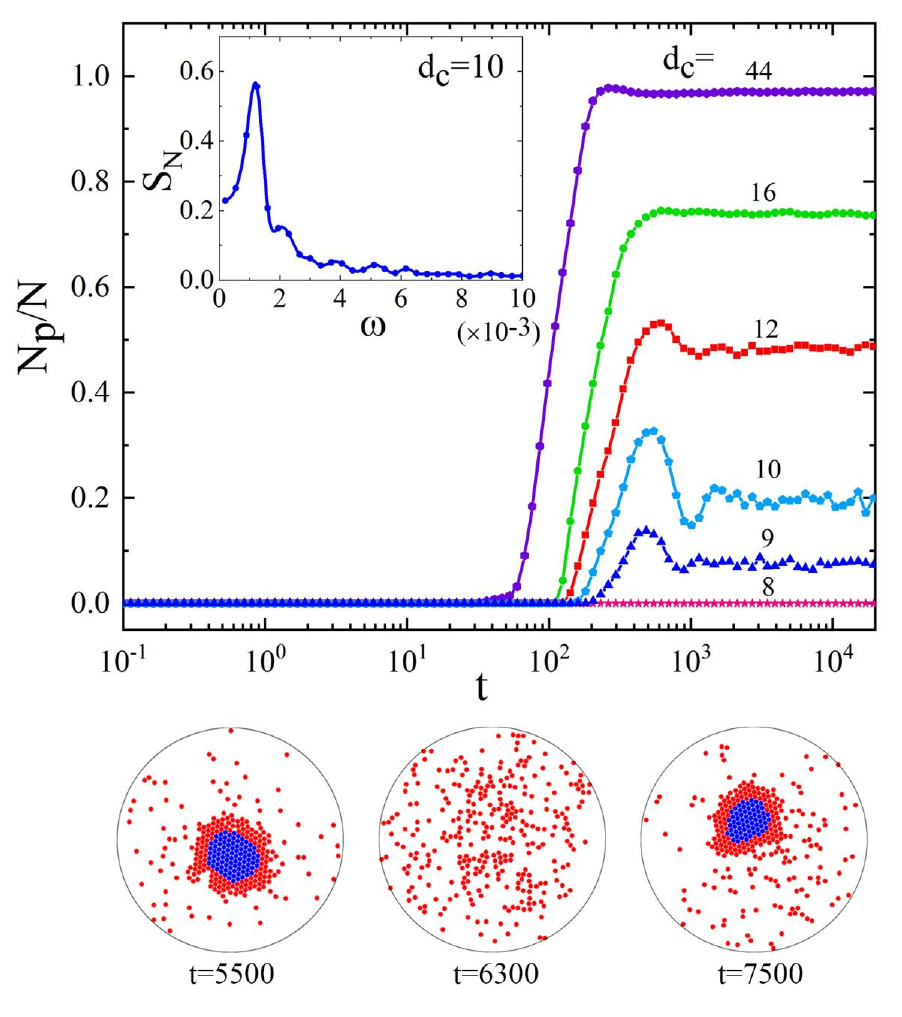}
\caption{Cluster stability. Top panel: the fraction of passive particles, $N_p(t)/N$, vs. $t$ for different $d_c$ (see legend).  The scattering angle distribution here is $p(\phi)=\delta (\phi)$, which corresponds to the limit $\lambda \to \infty$ of Eq. (\ref{phi}). Inset: frequency spectrum of the subtracted function $[N_p(t)-N_p(\infty)]/N$ in the stationary regime for $t\in ( 2\times 10^3, 2\times 10^4)$. Bottom panel: snapshots showing the aggregation and evaporation of an hexatic cluster for $d_c=10$. In both panels, active/passive particles are denoted by red/blue circles and the remaining simulation parameters are: $\alpha=\pi$, $D_\theta=0.001$, $v_0=0.5$, $R=45$, $r_0=1$, $N=304$, and $D_0=0.01$.  \label{F3}}
\end{figure}

\section {Results} \label{results}

The number of tunable
parameters of our model is quite large. In our simulations we kept the
particle radius, $r_0$, and self-propulsion speed, $v_0$, fixed, which
amounted to setting space and time units. The particle number, $N$, and the
cavity radius, $R$, played no key role as long as the suspension packing
fraction, $\phi_0=N(r_0/R)^2$, was kept sufficiently small (typically
$\phi_0<0.2$), to avoid steric clustering \cite{Fily}. We remind that the
active-passive transition threshold, $P_0(\alpha)$, of Eq. (\ref{Pth}) scales
like $N/R$. All remaining parameters, $D_0, D_\theta=1/\tau_\theta, \alpha,
d_c$ and $\lambda$, were varied to shed light on the underlying
collective dynamics.

Our main findings are summarized in Fig. \ref{F2} for the optimal case of
full visual perception, $\alpha=\pi$, persistence length,
$l_\theta=v_0\tau_\theta$ much larger than the cavity diameter, and small
translational noise, $D_0$ (whereby the time for a particle to diffuse a
distance of the order of its diameter is much larger than to self-propel the
same distance). The resulting 2D parameter space, $(\lambda, d_c)$, is
traversed by a continuous separatrix curve, $d_c$ vs. $\lambda$, whereby below (above) it
all JPs retain (lose) their active nature. Since the packing fraction of the simulated
active suspension is too small to trigger steric clustering, no active
aggregates were detected (region I). For small $\lambda$ values the suspension maintains
its initial uniform distribution, whereas at large $\lambda$,
transient, short-lived active clusters form and dissolve
(see Fig. 1 of the Supplemental Material \cite{SM})

Above the separatrix curve, the active-passive transitions induced by quorum
sensing, Eq. (\ref{Pth}), sustain the formation of large clusters, with core
consisting of passive particles. For large $d_c$ values, region II, one
typically observes a large passive condensate surrounded by a low-density gas
of active particles [see also the active and passive radial distributions
of Fig. \ref{F1}(d)], which bears a certain similarity with the situation
analyzed in Ref. \cite{Lozano}. The passive constituents of the condensate keep
fluctuating subject to thermal noise, as expected in a liquid phase. At large
$\lambda$, when the wall scatters the colliding particles mostly
toward the center of the cavity, the clustering mechanism exhibits additional
distinct features. Lowering $d_c$, we detected two more regions, III and IV,
characterized by very dense clusters made of a passive core surrounded by an
active layer; in both, the particles are closely packed into hexatic
structures \cite{hex}. The particles of cluster active layer show
larger motility than in the cluster core, but substantially lower than the
surrounding active gas particles, As a major difference,  in region III the clusters are stationary
in time, whereas in region IV the clusters appear and disappear over time.
The time oscillating clustering process of region IV is further analyzed in
Fig. \ref{F3}. In the top panel there we plotted the fraction of passive particles,
$N_p/N$, versus time for $\lambda \to \infty$. One notices immediately that for
$d_c$ values corresponding to the regions I-III, this ratio approaches a
steady state value after a transient of the order of the ballistic cavity
crossing time, $R/v_0$. In passing, we also remark for all curves
$N_p(t\to \infty)/N<1$, no matter what $d_c$,
which suggests that a gas of active particles is always at work.
Vice versa, as anticipated above, for system configurations in the
region IV, $N_p/N$ appears to execute persistent irregular time oscillations.
A spectral analysis of the time dependent ratio $N_p(t)/N$ confirms that: (i)
$N_p(t)$ fluctuates around a stationary asymptotic value $N_p(\infty) <1$; (ii)
the spectral density of the subtracted ratio, $[N_p(t)-N_p(\infty)]/N$ (an
example is shown in the inset of Fig. \ref{F3}), peaks around a finite frequency
of the order of $D_\theta$. We further observed that, on increasing $D_\theta$, region IV
in Fig. \ref{F2} shrinks and finally disappears.

The numerical results of Figs. \ref{F2} and \ref{F3} demonstrate the role of the boundary
in the cluster formation. At large $\lambda$, the distribution of the
scattering angle, $p(\phi)$, is strongly peaked around $\phi=0$; the boundary
exerts a lensing effect on the active particles by re-directing them toward
the center of the cavity. This clearly enhances the probability that the
sensing function, $P_i(\alpha)$, of the particles there overcome the threshold
value, $P_0(\alpha)$, of Eq. (\ref{Pth}), thus triggering the clustering
process. This is the key ``herding'' function of the gas of active particles continuously
bouncing between the cavity and the cluster border. Accordingly,
we noticed that in region II the clusters tend to be denser along the border.
In the opposite limit of wide scattering angle distribution, i.e., for
small $\lambda$, the active suspension is no longer focused toward the cavity
center and clustering is suppressed. The separatrix curve in Fig. \ref{F2}
clearly shows that for $\lambda \to 0$ clustering requires that $d_c \sim R$,
as implicit in the quorum sensing protocol adopted with Eqs. (\ref{Pa}) and
(\ref{Pth}).

\begin{figure}[tp]
\centering \includegraphics[width=6cm]{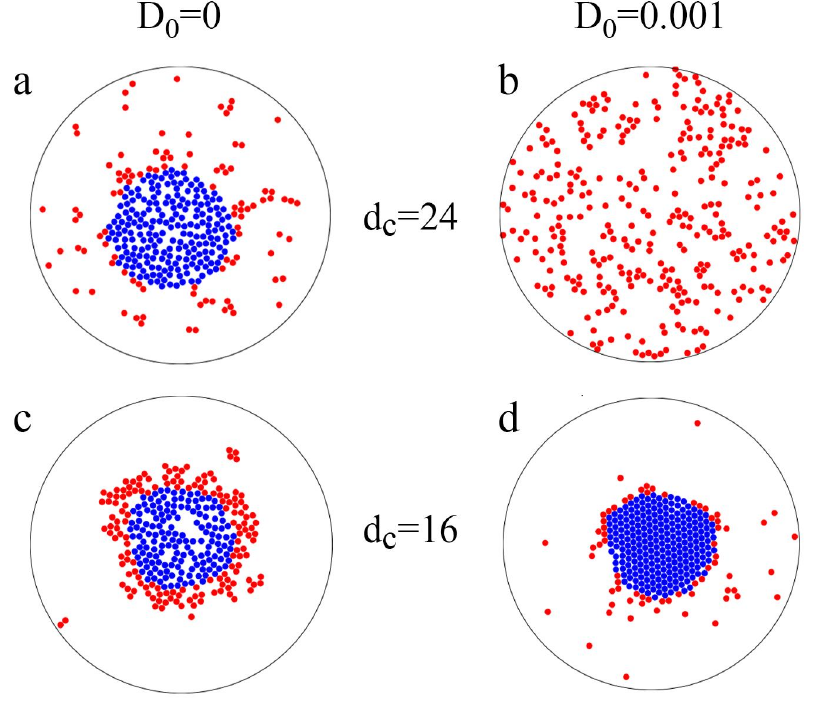}
\caption{Role of additive noise. Left panels: $D_0=0$ and snapshot time $t=10^5$; right panels: $D_0=0.001$ and $t=2\times 10^4$; top panels: $d_c=24$ and $D_\theta=0.01$; bottom panels: $d_c=16$ and $D_\theta=0.001$. The remaining simulation parameters are: $\alpha=\pi$, $v_0=0.5$, $R=45$, $r_0=1$, and $N=304$.  The $\phi$ distribution and the particle color code are as in Fig. \ref{F3}. \label{F4}}
\end{figure}

\section {Discussion and Conclusions} \label{conclusion}

The phase diagram of Fig. \ref{F2} was obtained for a convenient choice of
the tunable parameters $D_0, D_\theta$ and $\alpha$. We now briefly discuss
the role of these parameters in the clustering process.

{\em (i) role of spatial noise, $D_0$}. The additive noises $\xi_{x i}(t)$ and
$\xi_{yi}(t)$ in Eq. (\ref{LE}) keep the particles diffusing even after they
underwent the active-to-passive transition. This has a twofold consequence.
On one side, it hampers the cluster formation by delaying it in time and
pushing the separatrix curve of Fig. \ref{F2} to higher $d_c$ values [compare
Figs. \ref{F4}(a) and (b)]. On the other hand, for $D_0=0$ the passive
particles come immediately at rest after having reset their self-propulsion
velocity to zero. This leads to the buildup of frozen clusters with an amorphous glassy
structure [Fig. \ref{F4}(c)]. Vice versa, adding a little amount of noise
allows clusters to rearrange themselves in the denser hexatic structures of
region III [Fig. \ref{F4}(d)].

{\em (ii) role of rotational noise, $D_\theta$}. In Fig. \ref{F2} we assumed
that the particle persistence length, $l_\theta=v_0/D_\theta$, was much
larger than the cavity diameter. That choice was convenient in that it
enhanced the role of the boundary dynamics in the clustering process. Indeed,
under this condition, active JPs may hit the cavity walls repeatedly
before grouping at the center, where eventually undergo the active-to-passive
transition. To clarify the role of the persistence time,
$\tau_\theta=1/D_\theta$, we simulated the time evolution of the same
suspension for increasing values of $D_\theta$ and observed that cluster
formation gets, indeed, progressively suppressed (see the Supplemental
Material \cite{SM} for details). This comes as no surprise, since upon
increasing $D_\theta$, the persistence length, $l_\theta$, decreases and the
active particles' dynamics resembles more and more a standard Brownian motion with strength
$v_0^2/2D_\theta$.

\begin{figure}[tp]
\centering \includegraphics[width=7.5cm]{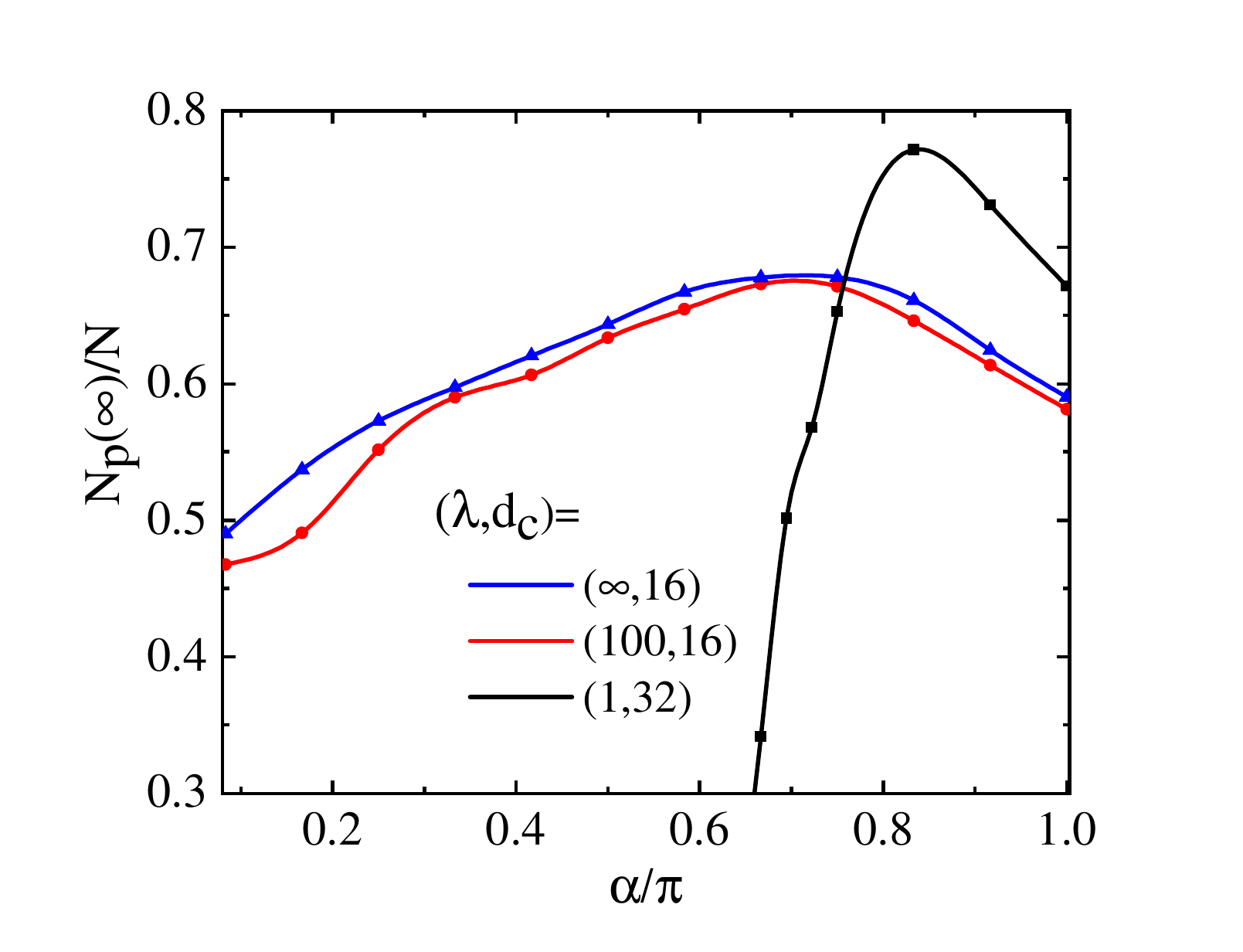}
\caption{Role of the visual angle, $\alpha$. Asymptotic value of $N_p(t)/N$ vs. $\alpha$ for different choices of the tunable parameters $(\lambda, d_c)$. The remaining simulation parameters are: $D_\theta=0.001$, $v_0=0.5$, $R=45$, $r_0=1$, $N=304$ and $D_0=0.05$.
\label{F5}}
\end{figure}

{\em (iii) role of the visual angle, $\alpha$}. We consider now cases when,
contrary to Fig. \ref{F2}, $\alpha < \pi$. This means that the neighbor
preception of particle $i$ is restricted to a visual cone directed along its
instantaneous self-propulsion velocity vector, ${\vec v_0}_i$
\cite{Lavergne2019}. This enhances the non-reciprocal nature of the particle
interactions.  As a consequence, the active-passive transitions at the
periphery of the forming clusters become more frequent. Indeed, an incoming
particle perceives a comparatively much larger neighbor density than a
particle moving outward. We remind here that all particles, active and
passive alike, keep rotating randomly  [third Eq. (\ref{LE})] with correlation time
$\tau_\theta$. This mechanism tends to destabilize the forming clusters, so
that one expects that shrinking the visual cone eventually suppresses
clustering. Our simulations confirm this guess, even though the asymptotic
value of the ratio $N_p(t)/N$ exhibits a non-monotonic $\alpha$ dependence
with a maximum for $\alpha/\pi \gtrsim (3/4)$ (Fig. \ref{F5}) -- compare
the snapshots for $d_c=16$ and $\alpha=(7/8)\pi$ in Fig. \ref{F1}(c) and $\alpha=\pi$ in Fig. \ref{F2}. 
We attribute his behavior to
the combined effect of the mechanism above and the $\alpha$ dependence of the
sensing threshold $P_0(\alpha)$ (see the Supplemental Material \cite{SM} for
details).

We conclude this report briefly discussing a limiting case, where the cavity
wall is replaced by periodic boundaries. We considered a square 2D simulation
box of size $L$: particle dynamics and quorum sensing protocols are the
same as for the circular cavity; as a difference, a particle $i$ crossing a box side, is
re-injected into the box through the opposite side with same self-propulsion
vector ${\vec v_0}_i$. In this regard, periodic boundaries are reminiscent of the
scattering wall of the initial model for $\lambda \to 0$, in that the
self-propulsion direction of the re-injected particle tends to be uniformly
distributed. Similarly to Fig. \ref{F2}, one then expects that
clustering only occurs for $d_0 \sim L/2$, as a consequence of the very
definition of the active-passive transition threshold, $P_0(\alpha)$.
Direct numerical simulations (not shown) confirm this expectation.

{\bf Acknowledgement}:
Y.L. is supported by the NSF China under grant  No. 12375037 and No. 11935010.

\end{document}